\documentclass[12pt,letter]{article}

\usepackage{graphicx, epsfig, color}
\def\eslt{\not\!\!{E_T}}
\def\to{\rightarrow}

\def\bi{\begin{itemize}}
\def\ei{\end{itemize}}

\def\ta{\tilde a}

\def\sps1ap{SPS1a$^\prime$}
\def\c1p{C1$^\prime$}

\def\tst{\tilde t}

\def\tg{\tilde g}

\def\tell{\tilde\ell}
\def\tq{\tilde q}
\def\tw{\widetilde W}
\def\tz{\widetilde Z}
\def\alt{\stackrel{<}{\sim}}
\def\agt{\stackrel{>}{\sim}}
\def\be{\begin{equation}}  
\def\ee{\end{equation}}  
\def\bea{\begin{eqnarray}}  
\def\eea{\end{eqnarray}}  
\def\beas{\begin{eqnarray*}}  
\def\eeas{\end{eqnarray*}}  
\newcommand\prd[3]{{\it Phys.\ Rev.\ }{\bf D #1} (#2) #3}

\newcommand\prl[3]{{\it Phys.\ Rev.\ Lett.\ }{\bf #1} (#2) #3}
\newcommand\plb[3]{{\it Phys.\ Lett.\ }{\bf B #1} (#2) #3}
\newcommand\jhep[3]{{\it J. High Energy Phys.\ }{\bf #1} (#2) #3}

\newcommand\npb[3]{{\it Nucl.\ Phys.\ }{\bf B #1} (#2) #3}

\newcommand\ptp[3]{{\it Prog.\ Theor.\ Phys.\ }{\bf #1} (#2) #3}

\newcommand{\hepph}[1]{hep-ph/#1}

\newcommand\ppnp[3]{{\it Prog.\ Part.\ Nucl.\ Phys.}{\bf  #1} (#2) #3}

\begin{document}
\begin{titlepage}

\vspace{0.5cm}
\begin{center}
{\Large \bf Direct and indirect detection of higgsino-like WIMPs:\\
concluding the story of electroweak naturalness
}\\ 
\vspace{1.2cm} \renewcommand{\thefootnote}{\fnsymbol{footnote}}
{\large Howard Baer$^1$\footnote[1]{Email: baer@nhn.ou.edu }, Vernon Barger$^2$\footnote[2]{Email: barger@pheno.wisc.edu } 
and Dan Mickelson$^1$\footnote[3]{Email: mickelso@nhn.ou.edu }
}\\ 
\vspace{1.2cm} \renewcommand{\thefootnote}{\arabic{footnote}}
{\it 
$^1$Dept. of Physics and Astronomy,
University of Oklahoma, Norman, OK 73019, USA \\
}
{\it 
$^2$Dept. of Physics,
University of Wisconsin, Madison, WI 53706, USA \\
}

\end{center}

\vspace{0.5cm}
\begin{abstract}
\noindent 
Supersymmetric models which fulfill the conditions of electroweak naturalness generally contain
light higgsinos with mass not too far from $m_h\simeq 125$ GeV, while other sparticles can be much heavier.
In $R$-parity conserving models, the lightest neutralino is then a higgsino-like WIMP (albeit with non-negligible
gaugino components), with thermal
relic density well below measured values. This leaves room for axions to function as
co-dark matter particles. The local WIMP abundance is then expected to be below standard estimates, 
and direct and indirect detection rates must be accordingly rescaled. We calculate rescaled 
direct and indirect higgsino-like WIMP detection rates in SUSY models that fulfill
the electroweak naturalness condition. In spite of the rescaling, we find that ton-scale noble liquid
detectors can probe the entire higgsino-like WIMP parameter space, so that these experiments
should either discover WIMPs or exclude the concept of electroweak naturalness in 
$R$-parity conserving natural SUSY models. 
Prospects for spin-dependent or indirect detection are more limited due in part to the
rescaling effect.
\vspace*{0.8cm}

\end{abstract}

\end{titlepage}

\section{Introduction}

The recent discovery of a Higgs-like resonance by Atlas and CMS collaborations\cite{atlas_h,cms_h} 
at $m_h\simeq 125$ GeV seemingly adds another supporting pillar to the theory of weak scale supersymmetry, since in the 
Minimal Supersymmetric Standard Model (MSSM)\cite{wss} we expect $m_h\simeq 115-135$ GeV\cite{mhiggs}.
Aside from solving the gauge hierarchy problem, previous supporting pillars arising from data 
include 1. the measured values of gauge couplings allow for
gauge coupling unification at a scale $Q\simeq 2\times 10^{16}$ GeV within the MSSM, 
2. the large value of the top quark mass is precisely what is needed to drive electroweak symmetry
breaking via radiative corrections and 3. SUSY is replete with several possible cold dark matter (CDM) 
candidates (neutralino/WIMP, gravitino, axino) in the form of the lightest SUSY particle (LSP).

Since a variety of SUSY model parameters enter into the scalar potential of the theory, and thus
contribute to the $W,\ Z$ and Higgs masses, it is widely expected that superpartners should exist
at or around the weak scale. This mantra has been repeated in numerous talks and papers over the past
decades, so we refer to it here as the story of SUSY electroweak naturalness\footnote{
We thank Yuri Gershtein for making this point}. Indeed, the concept of naturalness may
dictate to some degree when it is time to give up on weak scale supersymmetry should no signal 
be ultimately found\cite{feng}.

While the existence of a Higgs-like scalar at $\sim 125$ GeV is a boon for SUSY models, on the contrary, 
no signal for superpartners has yet emerged at LHC. 
In models such as the popular mSUGRA/CMSSM\cite{msugra}, 
the Atlas and CMS collaborations\cite{atlas_susy,cms_susy} now require
$m_{\tg}\agt 1.4$ TeV for $m_{\tq}\sim m_{\tg}$ and $m_{\tg}\agt 1$ TeV for $m_{\tq}\gg m_{\tg}$. 
Already this fact has led some astute physicists
to give up on weak scale SUSY\cite{shifman}, or to at least concede that weak scale SUSY is finetuned.
Thus, the recent LHC limits on sparticle masses seemingly exacerbate what is known as the
Little Hierarchy Problem (LHP): why is there such a disparity between the sparticle mass scale and the 
electroweak scale?

Before jumping to conclusions, it pays to scrutinize electroweak naturalness more closely.
We can be more precise if we re-phrase the LHP in the following terms: how is it that the $Z$-boson
mass can exist at just 91.2 GeV while gluino and squark masses are at, or even well beyond, 
the TeV scale? The answer proposed in Ref's \cite{ltr,sugra,rns} is that all the individual 
{\it weak scale} contributions feeding mass into $m_Z$ should be not too far from $m_Z$. 
The value of $m_Z$ in the MSSM is given by
\be
\frac{m_Z^2}{2} = \frac{(m_{H_d}^2+\Sigma_d^d)-(m_{H_u}^2+\Sigma_u^u)\tan^2\beta}{(\tan^2\beta -1)}
-\mu^2\simeq -m_{H_u}^2-\mu^2
\label{eq:mz},
\ee
where the latter approximate equality obtains for ratio-of-Higgs vevs
$\tan\beta\equiv v_u/v_d\agt 3$ and where the $\Sigma_{u}^{u}$  and $\Sigma_{d}^{d}$ terms represent 
the sum of various radiative corrections\cite{rns}.
To be quantitative, a finetuning measure 
\be
\Delta_{EW} =max_i\left| C_i/(m_Z^2/2)\ \right|
\ee
may be defined, where $C_i$ represents any of the terms on the right-hand-side of Eq.~\ref{eq:mz}
({\it e.g.} $C_{H_u}\equiv -m_{H_u}^2\tan^2\beta/(\tan^2\beta -1)$ and $C_\mu \equiv -\mu^2$).
The finetuning measure $\Delta_{EW}$ enjoys several advantages as discussed in Ref. \cite{rns}: 
it is 1. model-independent (within the MSSM) in that any model giving rise to the same weak scale
SUSY spectrum will have the same measure of finetuning, 2. conservative, 3. unambiguous, 4. predictive, 
5. falsifiable and 6. simple to encode. 

In the 2-parameter non-universal Higgs model (NUHM2)\cite{nuhm2}, with parameter space given by
\be
m_0,\ m_{1/2},\ A_0,\ \tan\beta,\ \mu,\ m_A ,
\ee
scans have found that values of $\Delta_{EW}$ as low as $\sim 10$ (corresponding to $\Delta_{EW}^{-1}\sim 10\%$
electroweak finetuning) could be found\cite{ltr,rns}. To achieve low $\Delta_{EW}$, 
one needs {\it a}) $|\mu|\sim m_Z\sim 100-200$ GeV, {\it b}) $m_{H_u}\sim (1-2)m_0$ so that
$m_{H_u}^2$ is driven radiatively (via the large top-quark Yukawa coupling) 
to small but negative values at the weak scale and {\it c}) moderate values of $m_{1/2}\sim 0.3-1$ TeV.
Scalar masses may exist in the range 1-10 TeV (even higher if one allows for split generations)
although large trilinear soft breaking terms $A_0\sim 1.5 m_0$ are required. The large $A$-terms
yield large mixing in the top-squark sector which simultaneously softens the radiative
corrections $\Sigma_u^u(\tst_1)$ and $\Sigma_u^u(\tst_2)$, while lifting the value of
$m_h$ into the 125 GeV range\cite{ltr}. While the NUHM2 model admits values of $\Delta_{EW}$ even below 10, 
the mSUGRA/CMSSM model is considerably more tuned, where a minimum $\Delta_{EW}\sim 100$ has been found, 
although typically $\Delta_{EW}$ for mSUGRA is more like $10^3-10^4$.
Models with low $\Delta_{EW}\alt 30$ have been dubbed {\it radiative natural SUSY} (RNS) since
the low electroweak finetuning is radiatively driven.

The sparticle spectra found for RNS models with $\Delta_{EW}\alt 30$ is characterized by:
\bi
\item light higgsino-like $\tw_1^\pm$ and $\tz_{1,2}$ with masses $\sim 100-300$~GeV,
\item gluinos with mass $m_{\tg}\sim 1-4$~TeV,
\item top squarks with $m_{\tst_1}\sim 1-2$~TeV and $m_{\tst_2}\sim 2-5$~TeV,
\item first/second generation squarks and sleptons with masses
$m_{\tq,\tell}\sim 1-8$~TeV. The $m_{\tq,\tell}$ range can be pushed up to
20-30~TeV if non-universality of generations with $m_0(1,2)> m_0(3)$ is
allowed.
\ei
The RNS model with the above spectra yields branching fractions 
$BF(b\to s\gamma )$ and $BF(B_s\to\mu^+\mu^-)$ in accord with measured values, unlike
many models with lighter top squarks\cite{nat}.

As far as testability goes, RNS models should yield observable signals from gluino
pair production at LHC14 with $\sim 300$ fb$^{-1}$ for $m_{\tg}\alt 1.6$ TeV\cite{lhcreach}. 
The LHC14
reach for SS diboson production from $pp\to\tw_2^\pm\tz_4\to W^\pm W^\pm +\eslt$ gives a reach
in terms of $m_{\tg}$ of $m_{\tg}\sim 1.8$ TeV for the same integrated luminosity\cite{lhcltr}. 
Since $m_{\tg}$ can range up to $\sim 4$ TeV for the RNS models, 
then LHC14 will not be able to definitively discover/exclude RNS.

Alternatively, the hallmark distinction of RNS models is the presence of four light higgsinos with 
masses $\sim 100-300$ GeV. A linear $e^+e^-$ collider operating at $\sqrt{s}\agt 2|\mu |$ 
can definitively discover/exclude RNS models, {\it e.g.} $\sqrt{s}\sim 0.5$ (1) TeV would 
access all models with $\Delta_{EW}\alt 25$ (70). Since the timescale for operation of a TeV-scale ILC
is of order 10-20 years, here we address instead the possibility of much earlier discovery of the higgsino-like
WIMPs expected from RNS models.
 
\section{Direct and indirect higgsino detection}

\begin{figure}[tbp]
\includegraphics[height=0.5\textheight]{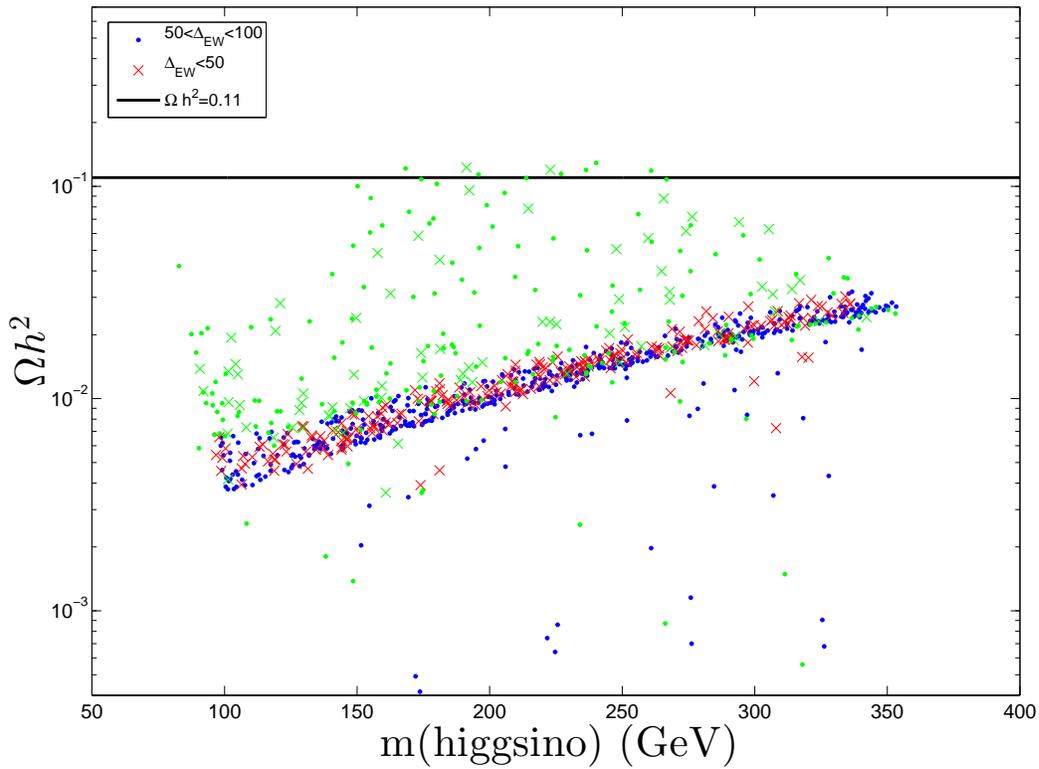}
\caption{Plot of standard thermal neutralino abundance 
$\Omega_{\tz_1}^{std}h^2$ versus $m(higgsino)$ 
from a scan over NUHM2 parameter space with $\Delta_{EW}<50$ (red crosses) and $\Delta_{EW}<100$ (blue dots). 
Green points are excluded by current direct/indirect WIMP search experiments.
We also show the central value of $\Omega_{CDM}h^2$ from WMAP9.
\label{fig:oh2}}
\end{figure}

In this paper, we generate RNS models from the Isasugra spectrum generator\cite{isasugra} 
using the same NUHM2 parameter space scan as utilized
in Ref. \cite{rns}. We will require that the model generated from each parameter set 
obeys the LEP2 bound that $m_{\tw_1}>103.5$ GeV, and will focus on models with 
$\Delta_{EW}<50$ (100) corresponding to better than 2\% (1\%) electroweak finetuning.
We calculate the ``standard thermal neutralino abundance'' $\Omega_{\tz_1}^{std}h^2$ using the 
IsaReD\cite{isared} relic density subroutine.\footnote{The standard thermal abundance refers 
to the neutralino density derived from assuming neutralinos in thermal equilibrium at high temperature 
followed by freeze-out due to the Universe's expansion.}
 We will accept only models with 
$\Omega_{\tz_1}^{std}h^2<0.12$ for reasons to be made clear shortly.
The relic abundance from RNS models is shown in Fig. \ref{fig:oh2}. The red crosses
have $\Delta_{EW}<50$ whilst blue dots have $\Delta_{EW}<100$. 
Green points we will find later are already excluded by direct/indirect WIMP search limits.
From the figure, 
we see a high density band extending from $\Omega_{\tz_1}^{std}h^2\sim 0.004$ for
$m_{\tz_1}\sim 100$ GeV to $\Omega_{\tz_1}^{std}h^2\sim 0.02$ for $m_{\tz_1}\sim 300$ GeV, 
{\it i.e.} there is typically a standard {\it underabundance} of higgsino dark matter compared to 
measurement from WMAP9\cite{wmap9} by a factor ranging from 3-25. There is some spread in these values above and below 
the main band from cases where $\mu$ is quite large and $m_{1/2}$ is small so that one has
instead a mixed higgsino-bino LSP state. 
The bulk of points above the band are already excluded as we shall see.
Thus, the mainly higgsinolike neutralino
by itself does not make a good CDM candidate. Additional new physics is needed to match the measured
dark matter density.

One compelling way to address the dark matter deficiency is by invoking the Peccei-Quinn-Weinberg-Wilczek
solution to the strong CP problem\cite{pqww} via introduction of a PQ symmetry and concommitant {\it axions}.
Since we are working in SUSY models, the axion will be accompanied by $R$-parity-even spin-0 saxions $s$
and $R$-parity-odd spin-$1\over 2$ axinos $\ta$\cite{axino_rev}. In gravity-mediation models (as assumed here),
the saxion and axino are expected to have masses $m_s\sim m_{\ta}\sim m_{3/2}\sim 5-20$ TeV\cite{axmass}, the same
mass scale as matter scalars $m_0$ in the theory. In this case, the dark matter would consist
of both axions and higgsinos acting as co-dark matter particles.

The relic abundance of mixed axion-neutralino CDM has been addressed in Ref's~\cite{ckls,blrs,bls,bbl}.
In \cite{bbl}, it was found that SUSY models with a standard overabundance of dark matter
are {\it still excluded} in the PQMSSM by a combination of dark matter overabundance 
constraints, BBN constraints and dark radiation constraints. However, SUSY models with a
standard underabundance of neutralinos are still allowed over large ranges of PQMSSM parameters.
For models with a standard underabundance, then thermal production and decay of axinos in the
early universe augments the neutralino abundance, sometimes by too much, other times not enough:
the former case would be excluded whilst in the latter case, 
the remaining abundance is made up of relic axions produced through
the usual vacuum misalignment mechanism. In addition, for large values of PQ breaking scale 
$f_a\sim 10^{13}-10^{16}$ GeV, then saxions can be produced at large rates via coherent oscillations.
The saxions can augment the neutralino abundance by decaying to SUSY particles ({\it e.g.} $s\to\tg\tg$),
or they can dilute all relics if they decay after freezeout into SM particles. Saxions may also
decay into axion pairs $s\to aa$ leading to production of dark radiation\cite{dark}. From the PQMSSM
scans in Ref. \cite{bbl}, it is found that in the case where $s\to aa$ branching fraction is at 
all substantial, then entropy dilution is {\it always} accompanied by a violation of either
BBN or dark radiation constraints (parametrized by $\Delta N_{eff}\agt 1.6$ where $\Delta N_{eff}$ is the
effective number of additional neutrinos beyond the SM value).
Thus, the scans over PQMSSM
parameter space in Ref. \cite{bbl} find that the standard underabundance can be augmented
by any factor leading to $\Omega_{\tz_1}^{std}h^2<\Omega_{\tz_1}h^2<0.11$, but not diminished 
without violating BBN or DR constraints. Models with too
much CDM production, or models which violate BBN or dark radiation constraints would be excluded.
The upshot is that in RNS models, for any particular parameter set, we expect the relic higgsino
abundance to lie somewhere between the standard value $\Omega_{\tz_1}^{std}h^2$ (which would correspond
to axion domination) up to the measured value $0.11$, in which case CDM would be higgsino-dominated.
The question then arises: what are the prospects for direct/indirect detection of relic higgsinos
in WIMP detection experiments?

\begin{figure}[tbp]
\includegraphics[height=0.5\textheight]{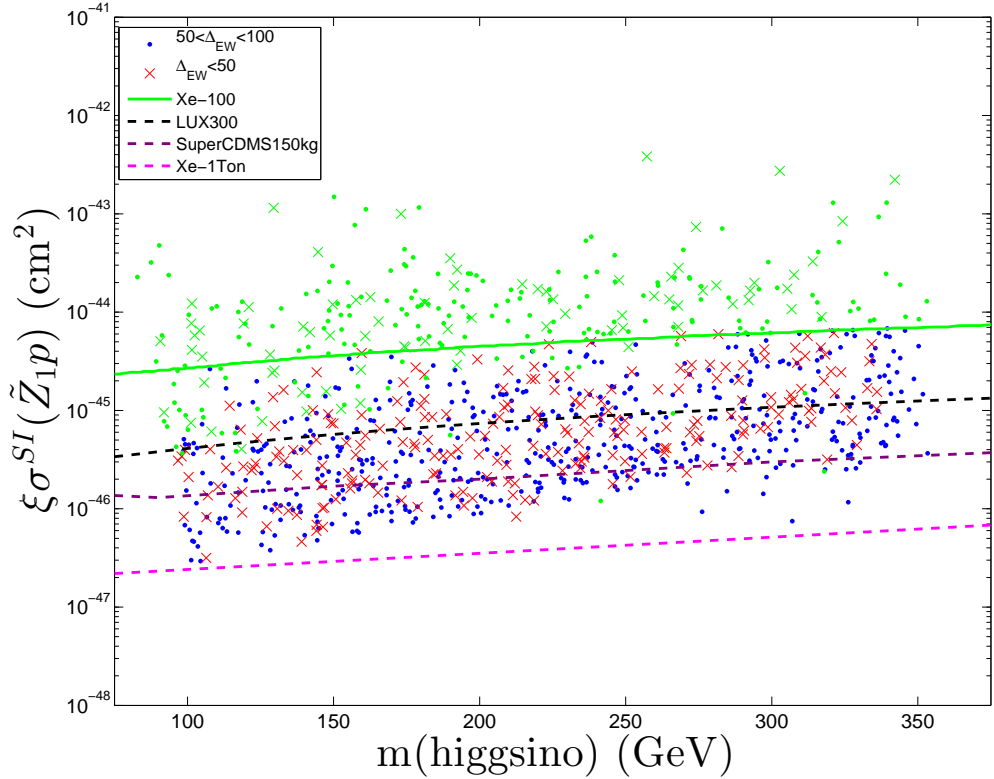}
\caption{Plot of rescaled higgsino-like WIMP spin-independent 
direct detection rate $\xi \sigma^{SI}(\tz_1 p)$ 
versus $m(higgsino)$ from a scan over NUHM2 parameter space with $\Delta_{EW}<50$ (red crosses)
and $\Delta_{EW}<100$ (blue dots). 
Green points are excluded by current direct/indirect WIMP search experiments.
We also show the current reach from $Xe$-100 experiment, 
and projected reaches of LUX, SuperCDMS 150 kg and $Xe$-1 ton.
\label{fig:SI}}
\end{figure}

In Fig.~\ref{fig:SI}, we show the spin-independent higgsino-proton scattering rate in $pb$
as calculated using IsaReS\cite{IsaReS}. The result is rescaled by a factor $\xi=\Omega_{\tz_1}^{std}h^2/0.11$
to account for the fact that the local relic abundance might be far less than the usually assumed value
$\rho_{local}\simeq 0.3$ GeV/cm$^3$, as suggested long ago by Bottino {\it et al.}\cite{bottino} 
(the remainder would be composed of axions). 
The higgsino-like WIMP in our case scatters from quarks and gluons mainly via $h$ exchange. 
The $\tz_1 -\tz_1 -h$ coupling involves a product of both higgsino and gaugino components. In the case of RNS models, 
the $\tz_1$ is mainly higgsino-like, but since $m_{1/2}$ is bounded from above by naturalness, the $\tz_1$
contains enough gaugino component that the coupling is never small:  in the notation of Ref. \cite{wss}
\be
{\cal L}\ni -X_{11}^h \overline{\tz}_1 \tz_1 h
\ee
where
\be
X_{11}^h =-{1\over 2}\left(v_2^{(1)}\sin\alpha -v_1^{(1)}\cos\alpha \right) 
\left(gv_3^{(1)}-g'v_4^{(1)}\right) ,
\ee
and where $v_1^{(1)}$ and $v_2^{(1)}$ are the higgsino components and $v_3^{(1)}$ and $v_4^{(1)}$ are
the gaugino components of the lightest neutralino, $\alpha$ is the Higgs mixing angle and $g$ and 
$g^\prime$ are $SU(2)_L$ and $U(1)_Y$ gauge couplings.
Thus, for SUSY models with low $\Delta_{EW}\alt 50-100$, the SI direct detection cross section is also 
bounded from below, even including the rescaling factor $\xi$.

From the Figure, we see that the current reach from 225 live-days
of Xe-100 running\cite{xe100} already bites into a significant spread of parameter points.
The excluded points are colored green.
The projected reach of the LUX 300 kg detector\cite{lux} is also shown by the black-dashed contour, which
should explore roughly half the allowed RNS points. 
The reach of SuperCDMS 150 kg detector\cite{cdms} is shown as the purple-dashed contour.
The projected reach of Xe-1-ton, a ton scale
liquid Xenon detector, is also shown\cite{xe1ton}. Our main result is this: the projected Xe-1-ton detector, 
or other comparable noble liquid detectors, can make a {\it complete}
exploration of the RNS parameter space. Since deployment of the Xe-1-ton detector is imminent,
it seems direct WIMP search experiments may either verify or exclude RNS models in the near future,
thus bringing the story of electroweak naturalness to a conclusion!

While the above result is indeed compelling, it is not a theorem, and is subject to several
reasonable assumptions. These include the assumption of $R$-parity conservation with a 
higgsino-like co-DM particle, along with the assumption of non-negligible $s\to aa$ decay rate.
If the $s\to aa$ decay rate is somehow forbidden or highly suppressed, then $s\to gg$ (gluons) 
may be dominant, 
in which case substantial entropy dilution can still occur\cite{blrs,bls}.

An alternative method for rescuing theories with an underabundance of WIMPs is to hypothesize
the existence of some scalar field (such as a modulus field from string theory) which can be 
produced via coherent oscillations, and which can engage in late decays injecting either more 
WIMPs (thus increasing the WIMP abundance) or entropy (thus decreasing the abundance\cite{mr}). 
The resulting abundance just depends on two parameters: 
the scalar field decay temperature  and the branching fraction into SUSY particles\cite{gg,endo,kane,dutta}. 
In our case, with a higgsino-like WIMP underabundance, the decaying scalar field
would increase the higgsino abundance to its measured value, and the rescaling factor $\xi =1$.
For this possibility, the results of Ref. \cite{rns} would apply.

\begin{figure}[tbp]
\includegraphics[height=0.5\textheight]{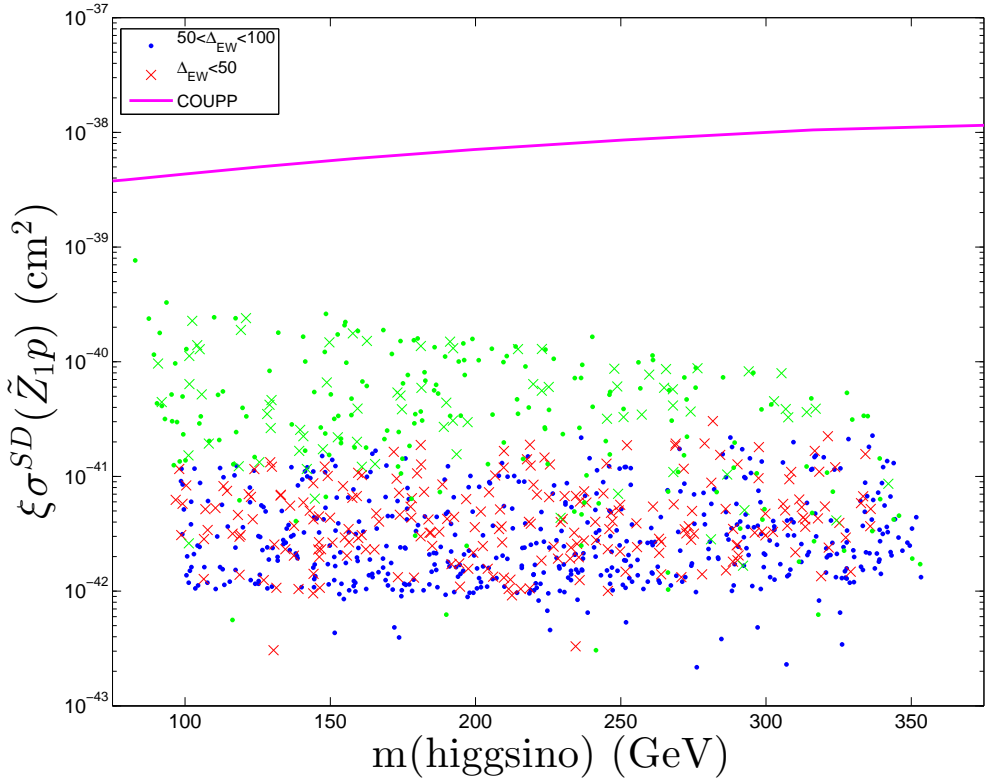}
\caption{Plot of rescaled spin-dependent higgsino-like WIMP detection rate $\xi \sigma^{SD}(\tz_1 p)$ 
versus $m(higgsino)$ from a scan over NUHM2 parameter space with $\Delta_{EW}<50$ (red crosses)
and $\Delta_{EW}<100$ (blue dots). 
Green points are excluded by current direct/indirect WIMP search experiments.
We also show current reach from the COUPP detector.
\label{fig:SD}}
\end{figure}

In Fig. \ref{fig:SD}, we show the rescaled {\it spin-dependent} higgsino-proton scattering
cross section $\xi\sigma^{SD}(\tz_1 p)$. Here we show recent limits from the COUPP\cite{coupp} 
detector. Current limits are still about an order of magnitude away from reaching the
predicted rates from RNS models.

\begin{figure}[tbp]
\includegraphics[height=0.5\textheight]{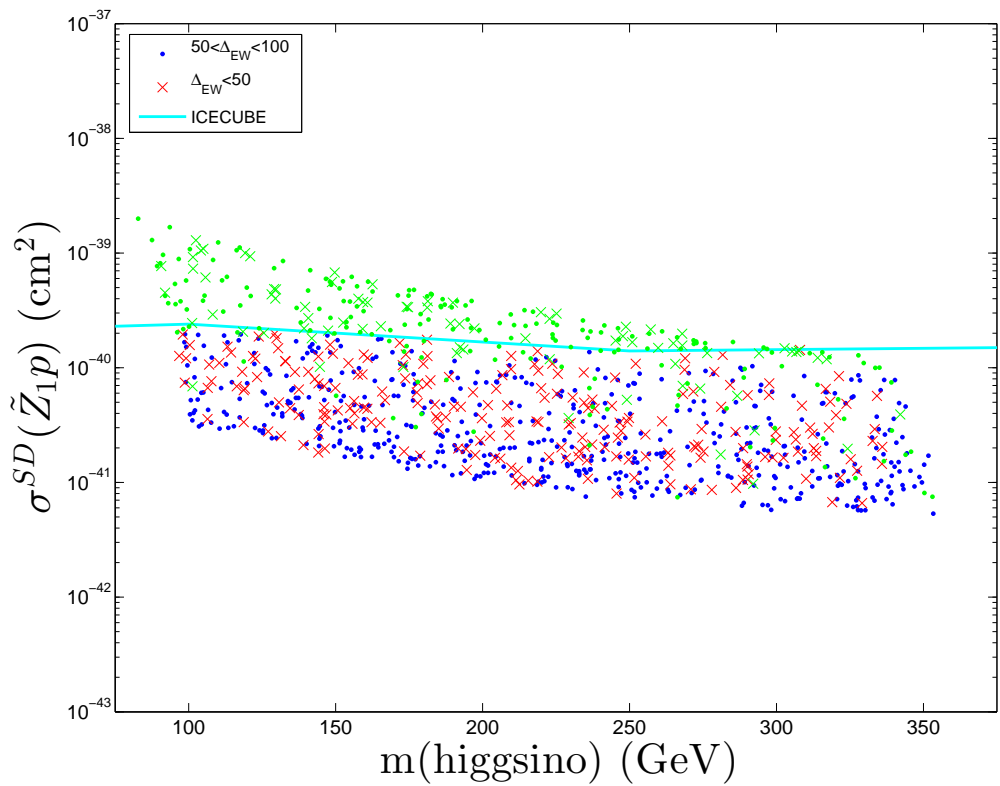}
\caption{Plot of (non-rescaled) spin-dependent higgsino-like WIMP detection rate $\sigma^{SD}(\tz_1 p)$ 
versus $m(higgsino)$ from a scan over NUHM2 parameter space with $\Delta_{EW}<50$ (red crosses)
and $\Delta_{EW}<100$ (blue dots). 
Green points are excluded by current direct/indirect WIMP search experiments.
We also show current reach from IceCube.
\label{fig:IC}}
\end{figure}

To compare against the current reach of IceCube\cite{icecube}, we show in Fig. \ref{fig:IC}
the value of $\sigma^{SD}(\tz_1 p)$, with no rescaling factor. 
Here,  the IceCube rates should not be rescaled since the IceCube detection depends on whether the Sun has 
equilibrated its core abundance between capture rate and annihilation rate\cite{gjk}. Typically for the Sun, 
equilibration is reached for almost all of SUSY parameter space\cite{bottino_nu}.
The IceCube limits have entered the RNS parameter space and excluded the largest values
of $\sigma^{SD}(\tz_1 p)$.  

\begin{figure}[tbp]
\includegraphics[height=0.5\textheight]{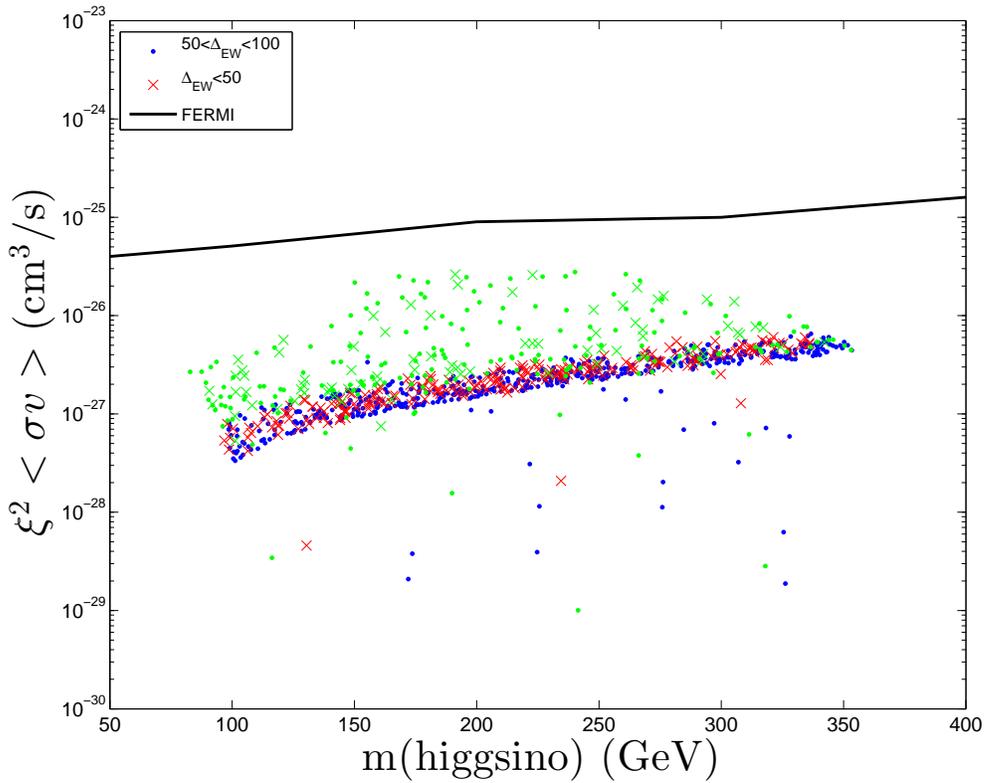}
\caption{Plot of rescaled $\xi^2 \langle\sigma v\rangle |_{v\to 0}$ 
versus $m(higgsino)$ from a scan over NUHM2 parameter space with $\Delta_{EW}<50$ (red crosses)
and $\Delta_{EW}<100$ (blue dots). 
Green points are excluded by current direct/indirect WIMP search experiments.
We also show current reach from Fermi LAT, 
Ref. \cite{fermi}.
\label{fig:sigv}}
\end{figure}

In Fig.~\ref{fig:sigv}, we show the rescaled thermally-averaged neutralino annihilation
cross section times relative velocity in the limit as $v\to 0$: $\xi^2\langle\sigma v\rangle|_{v\to 0}$.
This quantity enters into the rate expected from WIMP halo annihilations into
$\gamma$, $e^+$, $\bar{p}$ or $\bar{d}$. 
The rescaling appears as $\xi^2$ since limits depend on the square of the local WIMP abundance\cite{bottino_id}.
Anomalies in the positron and $\gamma$ spectra
have been reported, although the former may be attributed to pulsars\cite{pulsars}, 
while the latter 130 GeV gamma line may be instrumental. 
Soon to be released results from AMS-02 should clarify the
situation. On the plot, we show the limit derived from 
the Fermi LAT gamma ray observatory\cite{fermi} for WIMP annihilations into $WW$. 
These limits have not yet reached the RNS parameter space due in part to the squared rescaling factor.

\section{Conclusions:} 

In conclusion, we have found that SUSY models can elude the Little Hierarchy Problem
in the guise of radiative natural SUSY models which feature a mainly higgsino-like neutralino
that may act as a co-dark-matter particle along with the axion. While LHC can explore
a portion of RNS parameter space, an ILC can probe it entirely, although such a machine may be well over
a decade in the future. However, current WIMP direct detection experiments are biting into the meat 
of RNS parameter space, even if we take into account rescaling of the local abundance due 
to the fact that higgsinos may make up only a portion of the dark matter. 
LUX and ultimately SuperCDMS will probe further. The soon-to-be-deployed Xe-1-ton noble liquid detector 
should be able to {\it completely} explore the entire RNS parameter space (as shown in Fig. \ref{fig:SI}), 
thus either discovering a higgsino-like WIMP or rejecting the story of SUSY electroweak naturalness.
Complementary signals from spin-dependent and indirect WIMP detection channels 
are less likely since these are usually suppressed by
the reduced local WIMP density which is expected from models of mixed dark matter.

\section*{Acknowledgments}

We thank K. J. Bae, Y. Gershtein, P. Huang, A. Lessa, 
A. Mustafayev and X. Tata for discussions.
This work was supported in part by the US Department of Energy, Office of High
Energy Physics.

%

%
\end{document}